\documentclass[sn-aps, pdflatex, iicol]{sn-jnl}

\usepackage{showyourwork}
\usepackage{amsfonts,amssymb,amsmath}
\usepackage[nolist,nohyperlinks]{acronym}
\usepackage{bookmark}
\usepackage{xcolor}
\usepackage{array}
\usepackage{graphicx}
\usepackage{mathtools}
\usepackage{orcidlink}
\usepackage{tabularx}

\begin{document}

\title{Inferring cosmology from gravitational waves using non-parametric detector-frame mass distribution}

\author*[1]{Thomas~C.~K.~Ng\,\orcidlink{0000-0002-9491-1598}\,}\email{thomas.ng@link.cuhk.edu.hk}
\author[2,3]{Stefano~Rinaldi\,\orcidlink{0000-0001-5799-4155}\,}\email{stefano.rinaldi@uni-heidelberg.de}
\author[1]{Otto~A.~Hannuksela\,\orcidlink{0000-0002-3887-7137}\,}\email{hannuksela@phy.cuhk.edu.hk}

\affil[1]{Department of Physics, The Chinese University of Hong Kong, Shatin, Hong Kong}
\affil[2]{Institut~für~Theoretische~Astrophysik, ZAH, Universität~Heidelberg, Albert-Ueberle-Stra{\ss}e~2, 69120 Heidelberg, Germany}
\affil[3]{Dipartimento di Fisica e Astronomia ``G. Galilei'', Università di Padova, Via Marzolo 8, 35122 Padova, Italy}

\abstract{
    The challenge of understanding the Universe's dynamics, particularly the Hubble tension, requires precise measurements of the Hubble constant.
    Building upon the existing spectral-siren method, which capitalizes on population information from gravitational-wave sources, this paper explores an alternative method to analyze the population data to obtain the cosmological parameters in $\Lambda$CDM.
    We demonstrate how non-parametric methods, which are flexible models that can be used to agnostically reconstruct arbitrary probability densities, can be incorporated into this framework and leverage the detector-frame mass distribution to infer the cosmological parameters.
    We test our method with mock data and apply it to $70$ binary black hole mergers from the third gravitational-wave transient catalog of the LIGO-Virgo-KAGRA Collaboration.
}

\maketitle

\begin{acronym}
    \acro{GW}{gravitational-wave}
    \acro{LVK}{LIGO-Virgo-KAGRA Collaboration}
    \acro{PE}{parameter estimation}
    \acro{CMB}{cosmic microwave background}
    \acro{EM}{electromagnetic}
    \acro{(H)DPGMM}{hierarchy of Dirichlet process Gaussian mixture models}
    \acro{BBH}{binary-black-hole}
    \acro{FAR}{false alarm rate}
\end{acronym}

\section{Introduction}
\label{sec:introduction}

In recent years, the precision of cosmological measurements has improved significantly, leading to the discovery of the Hubble tension, a discrepancy between the value of the Hubble constant $H_0$ inferred from the \ac{CMB} \citep{Planck:2018vyg} and local measurements \citep{Riess:2021jrx}.
This tension has motivated the search for new methods to measure $H_0$ with high precision, and \ac{GW} sources detected by the \ac{LVK} \citep{KAGRA:2013rdx, LIGOScientific:2014pky, VIRGO:2014yos, KAGRA:2020tym} have emerged as a promising tool for this purpose \citep{LIGOScientific:2017adf, LIGOScientific:2021aug, Ezquiaga:2022zkx}.

To infer $H_0$ from \ac{GW} sources, one needs not only the luminosity distance which can be inferred from the signal but also the redshift of the source.
Some \ac{GW} sources can be associated with \ac{EM} counterparts, allowing for the measurement of the redshift, e.g., GW170817 \citep{LIGOScientific:2017adf, Guidorzi:2017ogy}.
However, most currently detectable \ac{GW} sources do not have \ac{EM} counterparts.
For these "dark sirens," one can assume \ac{GW} sources are in galaxies and associate the redshift of the galaxy to the \ac{GW} source, enabling an alternative way to measure cosmology \citep{Schutz:1986gp, DelPozzo:2011vcw, Gray:2019ksv, Mukherjee:2020hyn, Mukherjee:2022afz, Gray:2023wgj}.
Yet another method is to use physically motivated population models to infer $H_0$ by marginalizing over the redshift distribution of the sources.
This method is known as the spectral-siren method \citep{Farr:2019twy, You:2020wju, Mastrogiovanni:2021wsd, LIGOScientific:2021aug, Ezquiaga:2022zkx, Karathanasis:2022rtr}.

The fundamental idea of the spectral-siren method is to infer cosmology by comparing the observed mass population distribution with the intrinsic mass population distribution.
The observed distribution is redshifted due to the expansion of the Universe.
Cosmological parameters are inferred by relating the intrinsic mass distribution with the observed mass distribution.

The method works best when the intrinsic distribution is well understood.
However, the features present in the currently inferred mass distribution are not so well characterized yet, and their astrophysical origin is a debated topic \citep{Zevin:2017evb, Mapelli:2020vfa, Zevin:2020gbd, Mandel:2018hfr, Marchant:2023wno}.
In this case, the spectral-siren method depends on the choice of the population model, which can introduce biases in the inference \citep{Mukherjee:2021rtw, Mastrogiovanni:2021wsd, Pierra:2023deu, LIGOScientific:2020kqk, KAGRA:2021duu, LIGOScientific:2021aug}.
Furthermore, the intrinsic mass distribution is degenerate with cosmology.
For example, an intrinsic mass distribution with a linear redshift evolution is degenerate with $H_0$ at low redshift, as they can both redshift the observed distribution in the same way.
More complex redshift evolution can be degenerate with $H_0$ at high redshift.
While there is no reason to think a realistic intrinsic mass distribution will evolve linearly with redshift, this degeneracy still affects the inference of $H_0$.
Due to such degeneracies, astrophysics and cosmology are tightly linked to each other, and proper inference of the astrophysical and cosmological parameters should be done simultaneously, which makes the method computationally expensive.

A possible alternative approach, which we propose in this paper, is to use non-parametric methods to reconstruct the observed mass population distribution from the data.
Non-parametric methods allow us to reconstruct arbitrary probability densities without making assumptions about the form of the distribution \citep{Rinaldi:2021bhm}.
In our context, we can use non-parametric methods to reconstruct the observed mass population distribution directly from the data.
We compare the reconstructed distribution with the predicted distribution from some intrinsic population model and cosmological model, which allows us to infer the model parameters by minimizing the distance between the two distributions.

This approach allows us to extract the information contained in the observed data without making assumptions about the population model, cosmological model, or selection function.
At the same time, the reconstructed distribution can act as an intermediate observation-driven result that encodes both the intrinsic population information and the cosmological information.
Making use of the reconstructed distribution, we can easily explore the effect of using different models, as the reconstructed distribution can be reused for different models.
Furthermore, the intermediate result represents the information contained in the observed data without any assumptions about the population model, cosmological model, or selection function, which allows us to study the features of the observed population directly.

Other studies have made use of non-parametric methods for population studies \citep[e.g.,][]{Mandel:2016prl, Mandel:2018mve, Tiwari:2020vym, Rinaldi:2021bhm, Edelman:2021zkw, Sadiq:2021fin, Edelman:2022ydv, Callister:2023tgi, Ray:2023upk, Li:2023yyt, Farah:2024xub, MaganaHernandez:2024uty}, and now these methods are starting to be applied to the spectral siren method.
For example, \cite{Farah:2024xub, MaganaHernandez:2024uty, Li:2024rmi} independently used different non-parametric models to capture the features of the intrinsic mass distribution flexibly.

In Sec.~\ref{sec:method}, we describe our method in detail.
In Sec.~\ref{sec:mock_data}, we present the analysis setup and results of the mock data study.
In Sec.~\ref{sec:real_data}, we apply our method to the real data from the \ac{LVK}.
Finally, we conclude in Sec.~\ref{sec:conclusion} with a summary and outlook.

\section{Method}
\label{sec:method}

In this section, we present the framework we developed to infer both the intrinsic population model and the cosmological model simultaneously by leveraging the detector-frame mass population distribution.
We will make use of the notation summarized in Table~\ref{tab:notation}, taken from~\cite{Rinaldi:2021bhm, Rinaldi:2022kyg}.

\begin{table}
    \caption{Notation used in this paper}
    \begin{tabular}{cp{0.7\linewidth}}
        \toprule
        Notation & Description \\
        \midrule
        $m_1$ & Primary mass of the \ac{GW} sources in the source frame \\
        $q$ & Mass ratio of the \ac{GW} sources \\
        $z$ & Redshift of the \ac{GW} sources \\
        $d_L$ & Luminosity distance of the \ac{GW} sources \\
        $m^z_1$ & Primary mass of the \ac{BBH} sources in the detector frame \\
        $Y_t$ & detector-frame primary mass posterior probability distribution samples of the $t$-th \ac{GW} event \\
        $\mathbf{Y}$ & Set of detector-frame primary mass posterior probability distribution samples of all \ac{GW} events \\
        $\Theta_i$ & Hyperparameters of the \ac{(H)DPGMM} \\
        $\mathbf{\Theta}$ & Set of hyperparameters of the \ac{(H)DPGMM} \\
        $p(m^z_1|\mathbf{\Theta})$ & Detector-frame observed primary mass population distribution reconstructed with the \ac{(H)DPGMM} \\
        $\Lambda$ & Source-frame population model \\
        $\Omega$ & Cosmological model \\
        $p(m_1|\Lambda)$ & Source-frame primary mass population distribution assuming the population model $\Lambda$ \\
        $p(m^z_1|\Lambda, \Omega)$ & Detector-frame primary mass population distribution assuming the population model $\Lambda$ and cosmological model $\Omega$ \\
        $\mathrm{det}$ & detectability of the \ac{GW} events \\
        \botrule
    \end{tabular}
    \label{tab:notation}
\end{table}

Specifically, we propose two major modifications to the standard spectral-siren method.
First, instead of transforming the individual-event posterior probability distribution samples from the detector frame to the source frame during the \ac{PE} process, we transform the intrinsic population model to the detector frame before performing \ac{PE}.
Second, instead of directly computing the likelihood of the observed data given the intrinsic population model, we reconstruct the non-parametric observed population distribution from the data and compare it with the intrinsic population model.

With these modifications, we separate the spectral-siren method into two parts.
The first part is the reconstruction of the observed population from the data, which does not require any assumptions about the population model, and the intermediate result is completely data-driven.
We make use of the \ac{(H)DPGMM} developed in \cite{Rinaldi:2021bhm} as a non-parametric model to reconstruct the observed population distribution.
In the second part, we make use of all the other assumptions, namely the population model, cosmological model and selection function, to transform the source-frame population model into the detector frame.
This separation allows us to explore the impact of different assumptions easily.
The intermediate result of the first part can be reused as long as the data does not change, reducing the computational cost of the method.

Our method consists of three main steps: the non-parametric reconstruction of the detector-frame observed population, the transformation of the source-frame population model to the detector frame, and the remapping of the reconstructed distribution to the model parameters.

\subsection{Non-parametric reconstruction of detector-frame observed population}
\label{sec:reconstruction}

The first step of our method is to reconstruct the detector-frame observed population.
We choose to consider the primary mass $m_1$ of the \ac{BBH} sources, as we want to study the cosmological parameters with the spectral-siren method.
In principle, one could consider more parameters at once by reconstructing not only a single-parameter observed distribution but instead a multi-parameter observed distribution.
However, that would require a more sophisticated population model in the later steps, which is beyond the scope of this work.
Thus, we focus on the primary mass, which is one of the better astrophysically modelled quantities.

With the posterior probability distribution samples obtained from the \ac{PE} of each \ac{GW} event, we reconstruct the observed detector-frame primary mass population distribution.
We use the \ac{(H)DPGMM} developed in \cite{Rinaldi:2021bhm} as a non-parametric model to reconstruct the observed $m^z_1$ population distribution $p(m^z_1|\mathbf{\Theta})$.
In a nutshell, (H)DPGMM approximates the underlying probability density with a (potentially infinite) Gaussian mixture model,
\begin{equation}
    p(m_1^z) \simeq \sum_j^\infty w_j \mathcal{N}(m_1^z|\mu_j,\sigma_j)\,,
\end{equation}
where the relative weights of the components are inferred using a Dirichlet process.
In what follows, we will refer to the (H)DPGMM parameters collectively as $\Theta = \{\mathbf{w},\boldsymbol\mu,\boldsymbol\sigma\}$.
To perform the reconstruction, we use the package \textsc{figaro}\footnote{\textsc{figaro} is publicly available at \url{https://github.com/sterinaldi/FIGARO} and via \texttt{pip}.} \citep{Rinaldi:2024eep}.
Performing the reconstruction once will give a distribution $p(m^z_1|\Theta_i)$ that is consistent with $\mathbf{Y}$ with a given set of hyperparameters $\Theta_i$.
To account for the uncertainty in the reconstruction, we need to perform the reconstruction multiple times to obtain a set of distributions $p(m^z_1|\mathbf{\Theta})$, where $\mathbf{\Theta}$ is a set of $\Theta_i$.
The uncertainty of the reconstruction can then be quantified by the spread of the distributions.
It is important to note that the reconstruction is only valid in the region where the samples are present and cannot be extrapolated to regions where no samples are present.
In Fig.~\ref{fig:simulation_reconstruction}, we show an example of the reconstruction obtained with \textsc{figaro}.

This part of the method is completely data-driven.
Therefore, the intermediate result is independent of the population, cosmological model and selection function, and we can reuse the result for testing different models.
This reduces the computational cost of the method, as we only need to perform the reconstruction once to analyze different models.

\subsection{Transformation of source-frame population model}
\label{sec:transformation}

On the other hand, we assume a population model $\Lambda$ to obtain a source-frame primary mass $m_1$ population distribution $p(m_1|\Lambda)$, which represents the underlying population distribution given the model $\Lambda$.
Note that, either a parametric model can be used to only assume the form of the population distribution, or a fixed population model can be used to assume an exact population distribution.
We then transform $p(m_1|\Lambda)$ to the detector frame with a given cosmological model $\Omega$ to obtain the population distribution $p(m^z_1|\Lambda, \Omega)$ by
\begin{equation}
    \mathtoolsset{multlined-width=0.89\displaywidth}
    \begin{aligned}
        &p(m^z_1|\Lambda, \Omega, \mathrm{det}) \\
        &\begin{multlined}
            = \int p(m^z_1|m_1, z, \Lambda, \Omega, \mathrm{det}) \\ \times p(m_1, z|\Lambda, \Omega, \mathrm{det}) \mathrm{d}m_1 \mathrm{d}z
        \end{multlined} \\
        &\begin{multlined}
            = \int p(m^z_1|m_1, z) \\ \times \frac{p(\mathrm{det}|m_1, z, \Lambda, \Omega)p(m_1, z|\Lambda, \Omega)}{p(\mathrm{det}|\Lambda, \Omega)} \mathrm{d}m_1 \mathrm{d}z
        \end{multlined} \\
        &\begin{multlined}
            = \frac{1}{p(\mathrm{det}|\Lambda, \Omega)}\int \delta(m^z_1-m_1(1+z)) \\ \times p(\mathrm{det}|m_1, z, \Omega)p(m_1|\Lambda)p(z|\Lambda, \Omega) \mathrm{d}m_1 \mathrm{d}z
        \end{multlined}\\
        &\begin{multlined}
            \propto \int p\left(\mathrm{det}\middle|\frac{m^z_1}{1+z},z,\Omega\right) \\ \times p\left(\frac{m^z_1}{1+z}\middle|\Lambda\right)p(z|\Lambda, \Omega) \mathrm{d}z\,,
        \end{multlined}
    \end{aligned}
    \label{eq:transformation}
\end{equation}
where "$\mathrm{det}$" represents whether the event is detectable or not.
We assume the $m_1$ distribution does not evolve with $z$.

The selection effects are included in the derivation in Eq.~\eqref{eq:transformation} via the first term in the final expression, using the selection function.
In general, the selection function is a function of all the intrinsic binary parameters: the most important, however, are the component masses and the luminosity distance $d_L$.
In this work, since our intrinsic distribution includes only $m_1$ and $z$, we marginalized out the mass ratio dependence of the selection function assuming an intrinsic distribution $p(q) \propto q^{1.1}$.

We opt for reconstructing the observed distribution with the non-parametric method and constructing the intrinsic, detectable, detector-frame distribution $p(m^z_1|\Lambda, \Omega, \mathrm{det})$.
This is in contrast to the more standard procedure of agnostically reconstructing the intrinsic distribution and comparing it to our astrophysical model.
We do so to circumvent the limits of non-parametric methods concerning extrapolation required to account for selection effects.
If the selection function censors one area of the parameter space, such as the low-mass, high-redshift binaries, a data-driven method will not be able to characterize the distribution in that specific area, leading to diverging uncertainties in the reconstructed distribution.
Conversely, if we modelled the observed distribution and constructed the intrinsic distribution model non-parametrically, we would not be able to constrain the population behavior in the censored region.
This method of including selection effects in the analysis is an approximation of the real selection process, as pointed out in \cite{Essick:2023upv}: nonetheless, the large uncertainties currently associated with \ac{GW} measurements are most likely to be dominant with respect to the bias induced by this approximation.
Therefore, we include the selection function in the transformation, deferring the investigation of properly including selection effects in the reconstruction of the observed distribution in a future paper.

Qualitatively, the transformation of the intrinsic, detectable, detector-frame primary mass $p(m^z_1|\Lambda, \Omega, \mathrm{det})$ (Eq.~\eqref{eq:transformation}) redshifts the source-frame intrinsic population distribution to the detector frame, where the redshift is determined by the cosmological model.
The selection function is then applied to the redshifted distribution to obtain the intrinsic, detectable, detector-frame primary mass $p(m^z_1|\Lambda, \Omega, \mathrm{det})$.
With different $\Lambda$ or $\Omega$, the resulting $p(m^z_1|\Lambda, \Omega, \mathrm{det})$ will be different.
In Fig.~\ref{fig:simulation_transformation}, we show three different $p(m^z_1|\Lambda, \Omega, \mathrm{det})$ for three different assumed cosmologies.

In contrast to the first part of the method, which was entirely data-driven, the second part involving this transformation includes the assumptions about the population and cosmological model, as well as the selection function.
After obtaining $p(m^z_1|\Lambda, \Omega, \mathrm{det})$, we can perform the remapping to obtain the optimized $\Omega$ and $\Lambda$.

\subsection{Remapping to model parameters}
\label{sec:remapping}

With $p(m^z_1|\mathbf{\Theta})$ and $p(m^z_1|\Lambda, \Omega, \mathrm{det})$ obtained from Sec.~\ref{sec:reconstruction} and Sec.~\ref{sec:transformation} respectively, we need a way of translating each realization of the reconstructed distribution to a set of population and cosmological parameters.
To do so, we define a notion of "closeness" using the Jensen-Shannon distance $d_\mathrm{JS}$ between the two distributions.
$d_\mathrm{JS}$ is calculated as
\begin{equation}
    d_\mathrm{JS}(p, q) = \sqrt{\frac{D_\mathrm{KL}(p||m) + D_\mathrm{KL}(q||m)}{2}},
\end{equation}
where $p$ and $q$ are the two distributions to be compared, $m = (p + q) / 2$, and $D_\mathrm{KL}$ is the Kullback-Leibler divergence.
$d_\mathrm{JS}$ is a symmetric and smoothed version of $D_\mathrm{KL}$, and it is bounded between 0 and 1.

For each reconstructed distribution $p(m^z_1|\Theta_i)$, we find an optimized $\Lambda_i$ and $\Omega_i$ by minimizing $d_\mathrm{JS}$.
For the minimization, we use the \textsc{scipy} \citep{2020SciPy-NMeth} implementation of the modified Powell algorithm.
Note that we calculate $d_\mathrm{JS}$ over the range of $m^z_1$ values covered by the posterior probability distribution samples of all \ac{GW} events.
This is because the reconstruction is only valid in the region where observations are present, as mentioned in Sec.~\ref{sec:reconstruction}.
In Fig.~\ref{fig:simulation_comparison}, we show an illustration of the comparison between the two populations.

The resulting distribution of the optimized parameters is not a posterior probability distribution in a Bayesian sense, as we do not consider the likelihood of the observed data given the population model.
Instead, it is a distribution of the optimized parameters that are consistent with the observed data, distributed according to the uncertainty in the reconstruction of the population.
That is, the uncertainty in the comparison of the two populations is not considered in the resulting distribution.

The optimization procedure is highly efficient as it does not scale with the number of \ac{GW} events, since the optimization is performed for each reconstructed distribution independently.
Therefore, even as the number of \ac{GW} events increases, the runtime of the optimization remains constant, making our method scalable and robust for future large datasets.
This is advantageous for performing the analysis with different models when more \ac{GW} events are available, as the reconstruction can be reused.
Other than that, the optimization procedure can be performed in parallel for each reconstructed distribution, as each reconstructed distribution is independent.
This reduces the runtime as long as computational resources are available.

The analysis code we developed to implement the method presented in this section and the results of the following sections are publicly available on GitHub\footnote{\url{https://github.com/thomasckng/det-frame-cosmo-with-FIGARO/tree/main/src/scripts}} and Zenodo \citep{ng_2024_13968239} respectively.

\section{Mock data}
\label{sec:mock_data}

In this section, we test our method with a mock data study.
To prepare the mock data, we first generate %
  20000\unskip\label{output/n_true_samples.txt}\unskip%
 sets of $m_1$ and $z$ samples from a \textsc{power law + peak} distribution (see, e.g., Eq.~B4 in \cite{KAGRA:2021duu}) for the $m_1$ distribution and uniform distribution over comoving volume and source-frame time (see, e.g., Eq.~8 in \cite{KAGRA:2021duu}) for the $z$ distribution.
Here we assume the $m_1$ distribution does not evolve with $z$.
We then transform the $m_1$ and $z$ samples to $m^z_1$ and $d_L$ samples.
Here we choose the $\Lambda$CDM model with parameters consistent with the results in \cite{Planck:2018vyg} as the cosmological model for the transformation of the samples.

After the transformation, we apply a selection function to the $m^z_1$ and $d_L$ samples.
We use the physically modelled selection function implemented in \cite{Lorenzo-Medina:2024opt}.
We marginalize over $q$ in the selection function, to be consistent with the approach described in Sec.~\ref{sec:transformation}.
After applying the selection function, which filters out the non-detectable events based on their $m^z_1$ and $d_L$, we retained %
  309\unskip\label{output/n_obs_samples.txt}\unskip%
 sets of samples for further analysis.

We then generate the $m^z_1$ posterior probability distribution samples for each true $m^z_1$ sample by
\begin{gather*}
    I \sim \mathcal{N}(0, 0.03)\,,\\
    P \sim \mathcal{N}(\log(T_t)+I, 0.03)\,,\\
    Y_t = \exp(P)\,,
\end{gather*}
where $\mathcal{N}(\mu, \sigma)$ is a normal distribution with mean $\mu$ and standard deviation $\sigma$, $I$ is the difference between the true $m^z_1$ value and the measured mean in log space, $P$ is the posterior probability distribution sample in the log space, $T_t$ is the true $m^z_1$ sample, and $Y_t$ is the posterior probability distribution sample for the $t$-th \ac{GW} event.
For each true $m^z_1$ sample, we draw a random $I$ and generate $1000$ $Y_t$ samples.
This procedure ensures the posterior probability distribution samples are positive.
After preparing the mock data, we test our method with two cases: the inference of $H_0$ only and the inference of $H_0$ and a subset of the population model parameters.

\subsection{Inference of $H_0$}
\label{sec:inference_H0}

In this subsection, we first test our method with a simplified case where we only infer $H_0$ to demonstrate the basic idea of our method.
We assume the selection function, the true mass and redshift population model and its parameters, and the true cosmological model and its parameters except $H_0$ are known.

\begin{figure}
    \script{plot_simulation_reconstruction.py}
    \includegraphics[width=\linewidth]{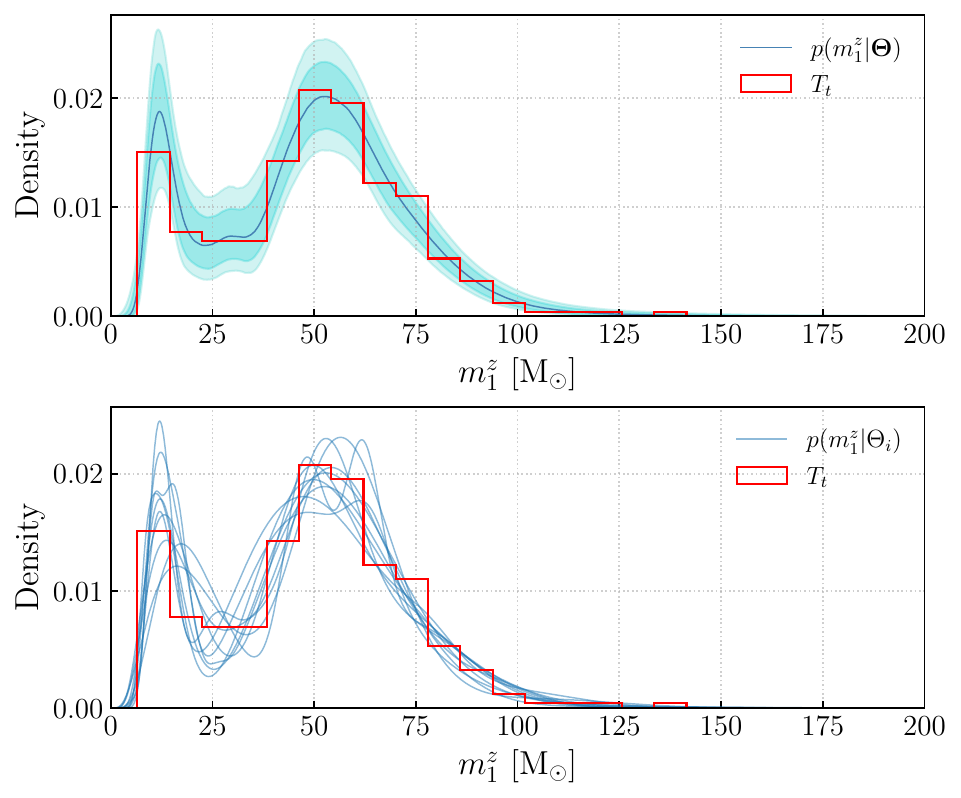}
    \caption{
        Top: non-parametric reconstruction of the observed distribution using 315 GW events (red histogram).
        The blue line marks the median of the reconstructed distributions, and the shaded regions represent the corresponding 68\% and 90\% credible intervals.
        Bottom: subset of individual draws for the non-parametric reconstruction.
    }
    \label{fig:simulation_reconstruction}
\end{figure}

We first reconstruct the observed detector-frame mass population distribution with \textsc{figaro} (see Sec.~\ref{sec:reconstruction}).
The result of the reconstruction is shown in Fig.~\ref{fig:simulation_reconstruction}.

\begin{figure}
    \script{plot_simulation_transformation.py}
    \includegraphics[width=\linewidth]{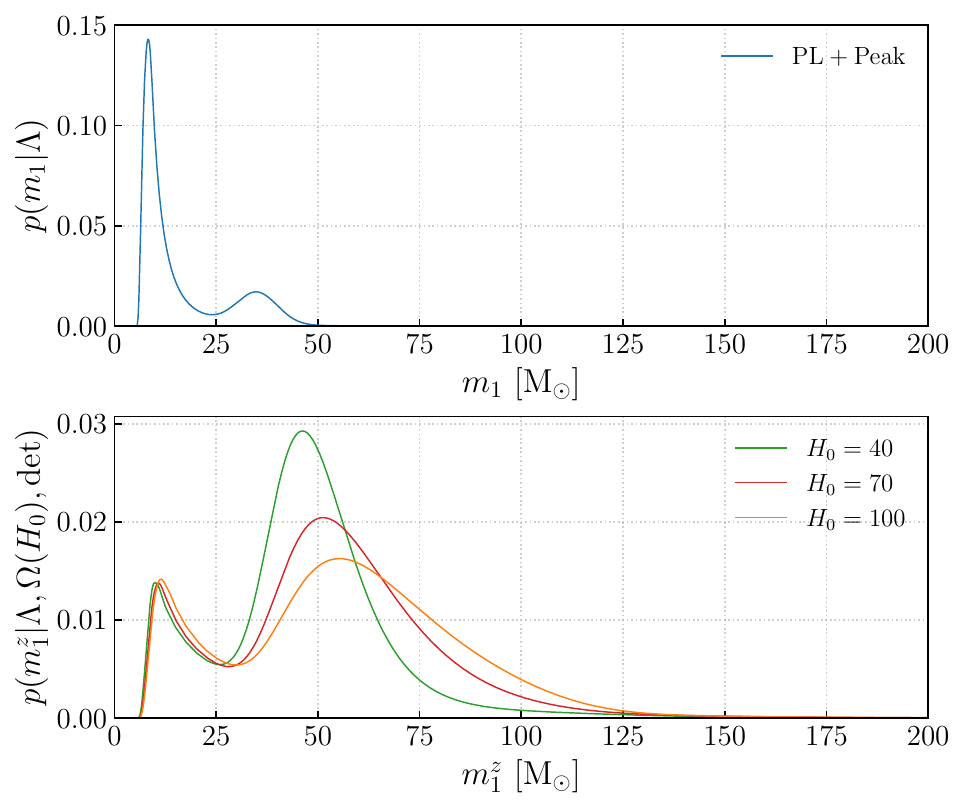}
    \caption{
        Top: source-frame $m_1$ population distribution.
        Bottom: detector-frame $m^z_1$ observed distribution for different $H_0$ values.
        The low-mass peak is suppressed due to the selection function.
    }
    \label{fig:simulation_transformation}
\end{figure}

Next, under the assumption that the population model, cosmological model and selection function are known, except for the value of $H_0$, we can perform the transformation mentioned in Sec.~\ref{sec:transformation}.
Following Eq.~\eqref{eq:transformation}, we can write the observed detector-frame mass population distribution as
\begin{equation}
    \mathtoolsset{multlined-width=0.89\displaywidth}
    \begin{aligned}
        &p(m^z_1|\Lambda, \Omega(H_0), \mathrm{det}) \\
        &\begin{multlined}
            \propto \int p\left(\mathrm{det}\middle|\frac{m^z_1}{1+z},z,\Omega(H_0)\right) \\ \times p\left(\frac{m^z_1}{1+z}\middle|\Lambda\right)p(z|\Lambda, \Omega(H_0)) \mathrm{d}z\,,
        \end{multlined}
    \end{aligned}
\end{equation}
where $\Omega(H_0)$ is the cosmological model with $H_0$ as a free parameter.
The result of the transformation is shown in Fig.~\ref{fig:simulation_transformation}.

\begin{figure}
    \script{plot_simulation_comparison.py}
    \includegraphics[width=\linewidth]{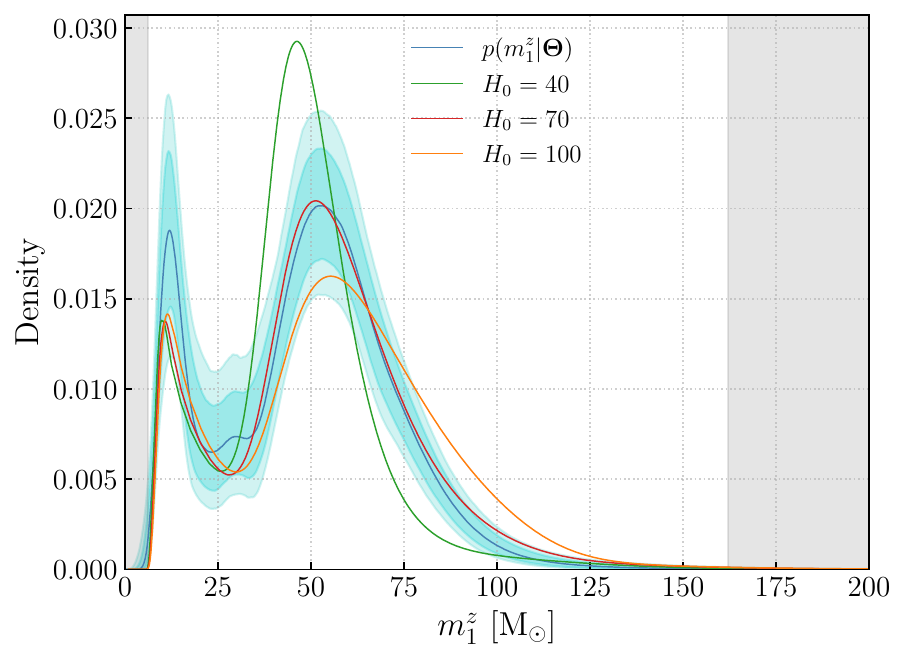}
    \caption{
        Comparison between the reconstructed observed distribution (blue line and shaded areas) and the predicted observed distribution with different $H_0$ choices.
        The gray shaded areas mark the boundaries for the $d_\mathrm{JS}$ calculation.
    }
    \label{fig:simulation_comparison}
\end{figure}

We then apply our optimization scheme to each reconstructed distribution to infer the optimized $H_0$.
In this case, since the parameter space is one-dimensional, instead of using the optimization algorithm, we rely on a grid-based search to find the optimized $H_0$ between $5$ and $150$ $\mathrm{km}\,\mathrm{s}^{-1}\,\mathrm{Mpc}^{-1}$.
For each $H_0$, we computed the $d_\mathrm{JS}$ between the two populations and found the optimized $H_0$ that minimizes $d_\mathrm{JS}$.
Figure~\ref{fig:simulation_comparison} shows an illustration of the comparison between the reconstructed observed population and the transformed population for different $H_0$.

\begin{figure}
    \script{plot_simulation_result_H0.py}
    \includegraphics[width=\linewidth]{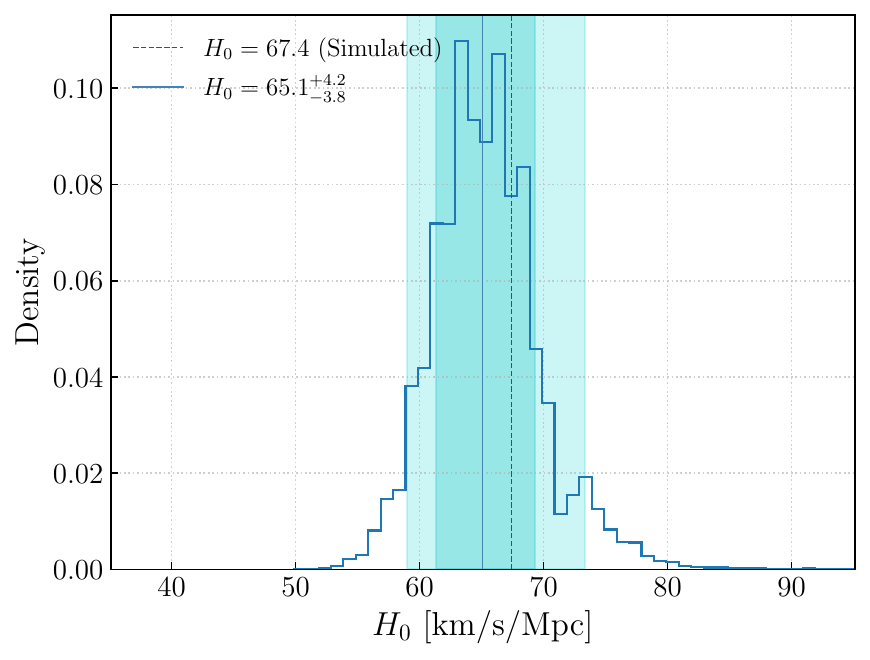}
    \caption{
        Optimized $H_0$ posterior distribution.
        The vertical blue line and shaded areas represent the median, 68\% and 90\% credible intervals respectively.
    }
    \label{fig:simulation_result_H0}
\end{figure}

The result of the remapping is shown in Fig.~\ref{fig:simulation_result_H0}.
The recovered $H_0$ distribution, whose uncertainty budget comes from the remapping of the individual-\ac{PE} uncertainty in the non-parametric reconstruction, is consistent with the simulated value.

\subsection{Inference of $H_0$ and $\Lambda$}
\label{sec:inference_multi}

In this subsection, we test our method with a more realistic case where we infer $H_0$ and a subset of the population model parameters.
Similar to the previous subsection, we assume perfect knowledge of both models, the selection function and all the parameters that are not objects of the inference.
Other than $H_0$, we include the power-law index $\alpha$, the peak mass $\mu$, and the peak width $\sigma$ of the \textsc{power law + peak} model as the free parameters in the analysis.
The minimum mass $m_\mathrm{min}$ and maximum mass $m_\mathrm{max}$, and the range of mass tapering at the low mass end $\delta$ are fixed.
These parameters, which control the boundary features of the observed population distribution, were chosen because the reconstructions are only valid in the region where the samples are present, and the reconstructed distributions may not be accurate at the boundaries depending on the boundary features of the observed population distribution.
In principle, more parameters can be inferred with our method, for example, the peak weight $w$ in the mass population model and the present mass density $\Omega_m$ in the $\Lambda$CDM model.
However, the optimization procedure becomes more computationally expensive as the number of parameters increases, therefore, we choose a subset of the parameters to demonstrate the idea of our method.
By performing the population inference together with the cosmological inference, we account for parameter degeneracies between the cosmological and population models.

The reconstruction of the observed $m^z_1$ population distribution is the same as in the previous subsection.
This also shows that the reconstruction is independent of the population model, cosmological model, and selection function, and we can reuse the result for different analysis setups.
As long as the events included in the analysis do not change, the reconstruction can be reused.
With more and more events in the near future, the standard spectral-siren method may become computationally expensive, and this feature of our method can become advantageous.

\begin{figure}
    \script{plot_simulation_result_multi.py}
    \includegraphics[width=\linewidth]{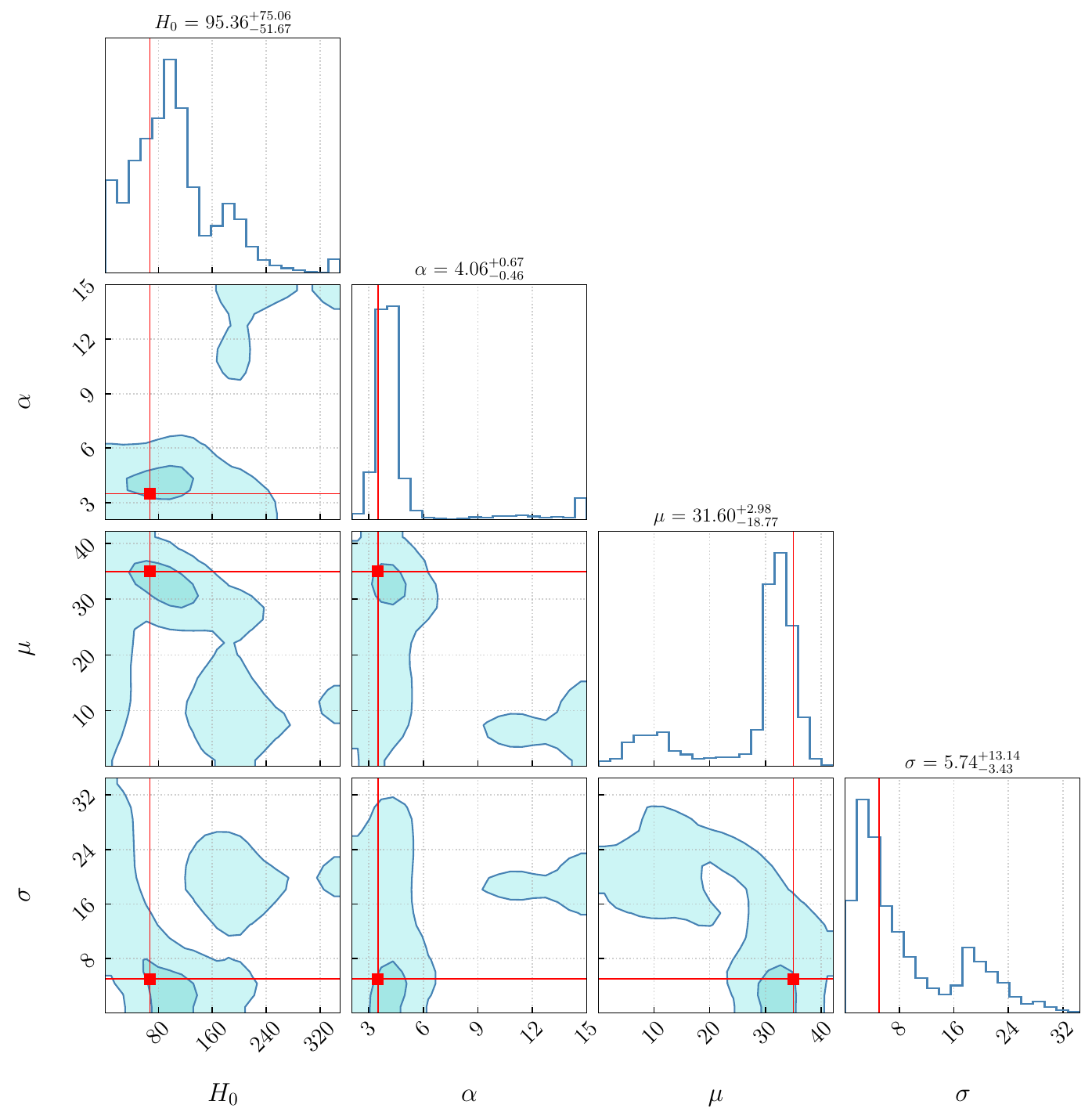}
    \caption{
        Optimized parameter distribution for the mock data inference.
        Simulated values are marked in red.
        The shaded regions represent the 39.35\% and 90\% credible intervals.
    }
    \label{fig:simulation_result_multi}
\end{figure}

In this example, instead of transforming the source-frame population model to the detector frame for each $H_0$ and performing a grid search, we define a function that calculates $d_\mathrm{JS}$ for a given set of the free parameters.
We then use the optimization algorithm mentioned in Sec.~\ref{sec:remapping} to find the optimized free parameters that minimize $d_\mathrm{JS}$.
The optimization method requires bounds for the free parameters, which we show in Table~\ref{tab:bounds}.
\begin{table}
    \caption{Free parameter bounds for the joint inference presented in Section~\ref{sec:inference_multi}.}
    \begin{tabularx}{\linewidth}{>{\centering\arraybackslash}X >{\centering\arraybackslash}X}
        \toprule
        Parameter & Bounds \\
        \midrule
        $H_0$ & $[1, 350]$ \\
        $\alpha$ & $[1.01, 15]$ \\
        $\mu$ & $[0.01, 70]$ \\
        $\sigma$ & $[0.01, 60]$ \\
        \botrule
    \end{tabularx}
    \label{tab:bounds}
\end{table}
The initial guesses for the optimization are chosen randomly from a uniform distribution within the bounds.
The result of the remapping is shown in Fig.~\ref{fig:simulation_result_multi}.
The recovered distribution, whose uncertainty budget comes from the remapping of the individual-\ac{PE} uncertainty in the non-parametric reconstruction, is consistent with the simulated values.

The consistency between the simulated and inferred parameters is expected, as the mock data is generated with the same population model and cosmological model we used in the analysis.
We can also see that the uncertainty in $H_0$ is larger compared to the analysis in Sec.~\ref{sec:inference_H0}, as more parameters are inferred simultaneously.

\section{Real data}
\label{sec:real_data}

In this section, we apply our method to the \ac{GW} events released by the \ac{LVK} as part of the GWTC-2.1 \cite{LIGOScientific:2021usb, ligo_scientific_collaboration_and_virgo_2022_6513631} and GWTC-3 \cite{KAGRA:2021vkt, ligo_scientific_collaboration_and_virgo_2023_8177023} catalogs.
We include all $70$ events labeled as BBH in Table~I of \cite{KAGRA:2021duu}, these correspond to all the events with $\mathrm{FAR_{min}} < 1\, \mathrm{yr}^{-1}$, where $\mathrm{FAR_{min}}$ is the smallest \ac{FAR} among all search pipelines.
The cut used to select these events is different from the one used in \cite{LIGOScientific:2021aug}, against which we will compare our results: there, the authors use a \ac{FAR} threshold of $\mathrm{FAR_{min}} < 0.25\, \mathrm{yr}^{-1}$, therefore the results we present here will be, in principle, different.
We opted for a larger \ac{FAR} threshold -- consistent, however, with the one used in population studies such as \cite{KAGRA:2021duu} -- to have a larger number of events to include in the non-parametric analysis.

\begin{figure}
    \script{plot_real_reconstruction.py}
    \includegraphics[width=\linewidth]{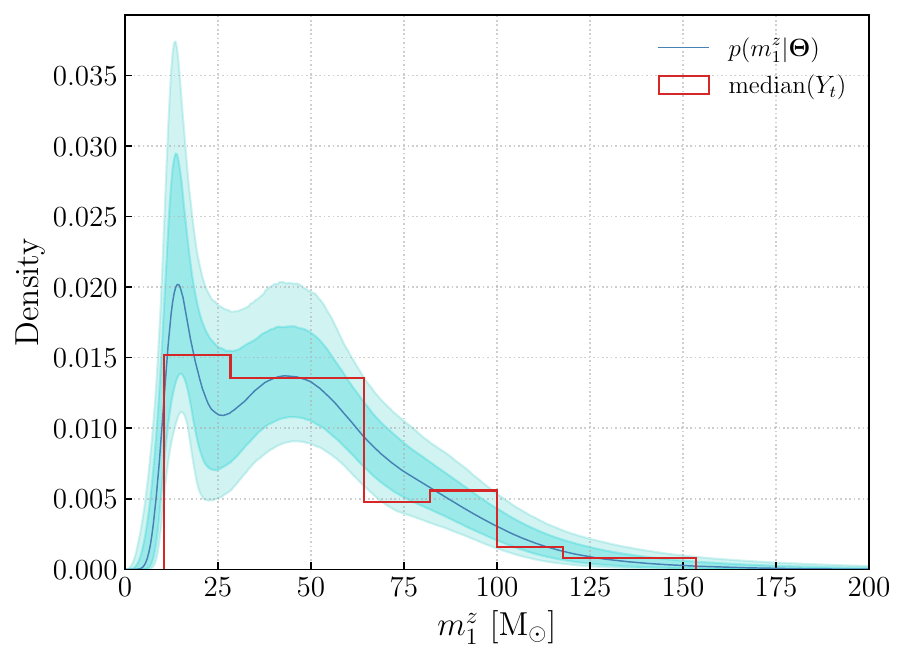}
    \caption{
        Median non-parametric reconstruction (blue line) and median values of the $m^z_1$ posterior probability distribution samples of the \ac{GW} events (red histogram).
        The shaded regions represent the 68\% and 90\% credible intervals for the non-parametric reconstruction.
    }
    \label{fig:real_reconstruction}
\end{figure}

Similar to Sec.~\ref{sec:mock_data}, we first reconstruct the observed $m^z_1$ population distribution.
We use the posterior probability distribution samples of the \ac{GW} events from \cite{LIGOScientific:2019lzm, KAGRA:2023pio}.
$10000$ samples are randomly drawn from the posterior probability distribution samples of each \ac{GW} event to perform the reconstruction.
The result of the reconstruction is shown in Fig.~\ref{fig:real_reconstruction}.

Next, we chose the \textsc{power law + peak} model as the source-frame population model $\Lambda$ and the $\Lambda$CDM model as the cosmological model $\Omega$.
For the redshift population model, instead of the uniform distribution over comoving volume and source-frame time, we used the \textsc{power law} redshift evolution model described in Eq.~8 of \cite{KAGRA:2021duu}.
Similar to Sec.~\ref{sec:inference_multi}, we fixed some parameters of the population model and the cosmological model.
For the cosmological model, we fixed the parameters to the results from \cite{Planck:2018vyg} except $H_0$.
For the population model, we fixed $\alpha$, $w$, $m_\mathrm{min}$, $m_\mathrm{max}$, and $\delta$ to the results from \cite{LIGOScientific:2021aug}.
Note that the fixed parameters are not necessarily the true values, but they can be considered as a reasonable guess for the parameters.
The fixed parameters of the population model are shown in Table~\ref{tab:fixed_parameters}.
\begin{table}
    \caption{Fixed parameters of the $\Lambda$CDM model (left) and the \textsc{power law + peak} model (right).}
    \begin{tabularx}{\linewidth}{>{\centering\arraybackslash}X >{\centering\arraybackslash}X | >{\centering\arraybackslash}X >{\centering\arraybackslash}X}
        \toprule
        Parameter & Value & Parameter & Value \\
        \midrule
         & & $w$ & $0.024$ \\
        $\Omega_m$ & $0.315$ & $\alpha$ & $4.2\ \mathrm{M}_\odot$ \\
        $\Omega_\Lambda$ & $0.685$ & $m_\mathrm{min}$ & $5\ \mathrm{M}_\odot$ \\
        $w$ & $-1$ & $m_\mathrm{max}$ & $109\ \mathrm{M}_\odot$ \\
         & & $\delta$ & $4.9\ \mathrm{M}_\odot$ \\
        \botrule
    \end{tabularx}
    \label{tab:fixed_parameters}
\end{table}

\begin{figure}
    \script{plot_real_result_multi.py}
    \includegraphics[width=\linewidth]{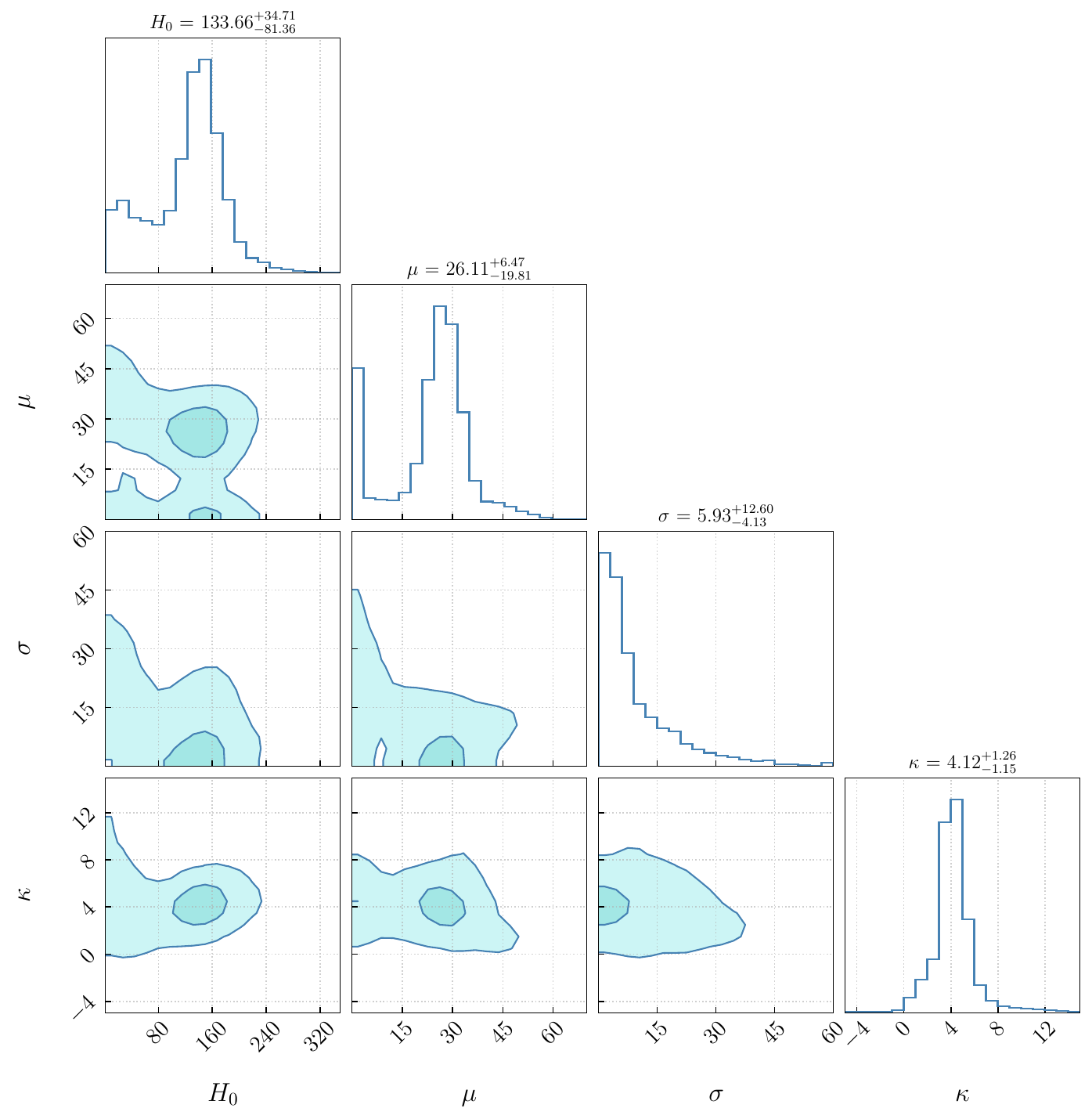}
    \caption{
        Optimized parameter distribution for the real data inference.
        The shaded regions represent the 39.35\% and 90\% credible intervals.
    }
    \label{fig:real_result_multi}
\end{figure}

We then perform the remaining steps in the same way as Sec.~\ref{sec:inference_multi}.
We set the bounds for the power-law index of the redshift population model $\kappa$ to $[-100, 100]$, and the bounds for other free parameters are the same as in Table~\ref{tab:bounds}.
The result of the remapping is shown in Fig.~\ref{fig:real_result_multi}, and it is consistent with the posterior distribution reported in \cite{LIGOScientific:2021aug}.
We find %
  $H_0 = 132.56^{+37.22}_{-92.15} \, \mathrm{km/s/Mpc}$\unskip\label{output/real_H0.txt}\unskip%
, (68\% credible interval around the median). Nonetheless, Figure~\ref{fig:real_result_multi} shows a peak in the marginalized posterior distribution of $H_0$ around $140\, \mathrm{km/s/Mpc}$, with a width of around $30-35\, \mathrm{km/s/Mpc}$ roughly consistent with the uncertainty reported in \cite{LIGOScientific:2021aug} ($H_0 = 50^{+37}_{-30} \, \mathrm{km/s/Mpc}$).

The reason for the peak differing from the one reported in \cite{LIGOScientific:2021aug}, which is around $50\, \mathrm{km/s/Mpc}$, can be attributed to two main factors.
First, the choice of the Jensen-Shannon distance as a measure of ``closeness'' between the non-parametric reconstruction and the predicted population is an arbitrary choice and, despite being a commonly used metric, does not stem from a specific derivation and/or assumption on the process that generated the observations.
As we stated in Section~\ref{sec:remapping}, the output of our method is not a posterior distribution in the strict sense due to this choice: the derivation of a well-grounded metric to connect non-parametric and parametric distributions will be discussed in a future paper.
Another possibility is the approximation made while including the selection effects: as discussed in \cite{Essick:2023upv}, this choice can lead, to some extent, to a bias in the reconstruction, and thus be the reason why the peak of the posterior distribution for $H_0$ is shifted.
Accounting for selection effects while reconstructing the observed distribution is in principle possible -- although not explored in the currently available literature -- and will be, as well, investigated in the future.
Despite these challenges, our method provides a foundation for exploring different models efficiently and can significantly aid in the analysis of large datasets expected in the near future.

\section{Conclusion}
\label{sec:conclusion}

In this paper, we proposed a method to infer both the intrinsic population model and the cosmological model simultaneously by leveraging the detector-frame mass population distribution and making use of non-parametric methods.
We demonstrated the method with a mock data study and applied it to the real data from the \ac{LVK}.
With the numbers of observed \ac{GW} events that are expected to grow larger and larger, potentially hitting the thousand mark by the end of the fifth observing run, performing a complete analysis from scratch with different astrophysical models will soon become too computationally expensive.
The method we propose in this paper can help mitigate this issue by providing a fast and efficient way to explore different models before performing the full hierarchical Bayesian inference, especially keeping in mind that, if a wrong population model is chosen, the inference result will be biased \citep{Mukherjee:2021rtw, Mastrogiovanni:2021wsd, Pierra:2023deu, LIGOScientific:2020kqk, KAGRA:2021duu, LIGOScientific:2021aug}.
This makes our method a powerful diagnostic tool for model selection.

To further improve the method, there are several directions we can explore.
First, as mentioned in Sec.~\ref{sec:remapping}, our method is not a Bayesian inference, as we only consider the optimization of the parameters.
To perform a full Bayesian inference, we need to develop a likelihood function that considers the comparison of the two populations.
With this likelihood, our method will give the same result as the standard hierarchical Bayesian inference, as long as the number of reconstructions is large enough.
This will allow us to perform efficient Bayesian model selection, as different models can be tested without repeating the reconstruction of the observed population distribution.
We will explore this possibility in future work.

Another potential improvement is to consider not only the primary mass distribution.
For example, we can consider the joint distribution of the primary mass, mass ratio, and redshift.
The reconstruction of the joint observed distribution can also be performed in the current framework.
However, we need a joint population model that considers all the parameters, including the possible correlations between the parameters.
Performing the analysis of the joint distribution will allow us to study mass population models that evolve with redshift.
Furthermore, the inference of cosmological parameters will be more precise, as the information from the luminosity distance is also considered.

\backmatter

\bmhead{Acknowledgements}

We thank Grégoire Pierra for helpful discussions.
T.~N. and S.~R. acknowledge financial support from the German Excellence Strategy via the Heidelberg Cluster of Excellence (EXC 2181 - 390900948) STRUCTURES.
S.~R. acknowledges financial support from the European Research Council for the ERC Consolidator grant DEMOBLACK, under contract no. 770017. 
We acknowledge the use of computing facilities supported by grants from the Croucher Innovation Award from the Croucher Foundation Hong Kong.
This research made use of the bwForCluster Helix: the authors acknowledge support by the state of Baden-Württemberg through bwHPC and the German Research Foundation (DFG) through grant INST 35/1597-1 FUGG.
This research has made use of data or software obtained from the Gravitational Wave Open Science Center (gwosc.org), a service of the LIGO Scientific Collaboration, the Virgo Collaboration, and KAGRA. This material is based upon work supported by NSF's LIGO Laboratory which is a major facility fully funded by the National Science Foundation, as well as the Science and Technology Facilities Council (STFC) of the United Kingdom, the Max-Planck-Society (MPS), and the State of Niedersachsen/Germany for support of the construction of Advanced LIGO and construction and operation of the GEO600 detector. Additional support for Advanced LIGO was provided by the Australian Research Council. Virgo is funded, through the European Gravitational Observatory (EGO), by the French Centre National de Recherche Scientifique (CNRS), the Italian Istituto Nazionale di Fisica Nucleare (INFN) and the Dutch Nikhef, with contributions by institutions from Belgium, Germany, Greece, Hungary, Ireland, Japan, Monaco, Poland, Portugal, Spain. KAGRA is supported by Ministry of Education, Culture, Sports, Science and Technology (MEXT), Japan Society for the Promotion of Science (JSPS) in Japan; National Research Foundation (NRF) and Ministry of Science and ICT (MSIT) in Korea; Academia Sinica (AS) and National Science and Technology Council (NSTC) in Taiwan.
This paper was compiled using \textsc{showyourwork} \cite{Luger2021} to facilitate reproducibility.

\bibliography{bib.bib}

\begin{thebibliography}{10}
\providecommand{\url}[1]{{#1}}
\providecommand{\urlprefix}{URL }
\providecommand{\doi}[1]{\url{https://doi.org/#1}}
\bibcommenthead

\bibitem{Planck:2018vyg}
N.~Aghanim, et~al., {Planck 2018 results. VI. Cosmological parameters}.
\newblock Astron. Astrophys. \textbf{641}, A6 (2020).
\newblock \doi{10.1051/0004-6361/201833910}.
\newblock [Erratum: Astron.Astrophys. 652, C4 (2021)].
\newblock {\href{https://arxiv.org/abs/1807.06209}{{arXiv:1807.06209}}}
  {[astro-ph.CO]}

\bibitem{Riess:2021jrx}
A.G. Riess, et~al., {A Comprehensive Measurement of the Local Value of the
  Hubble Constant with 1 km s$^{-1}$ Mpc$^{-1}$ Uncertainty from the Hubble
  Space Telescope and the SH0ES Team}.
\newblock Astrophys. J. Lett. \textbf{934}(1), L7 (2022).
\newblock \doi{10.3847/2041-8213/ac5c5b}.
\newblock {\href{https://arxiv.org/abs/2112.04510}{{arXiv:2112.04510}}}
  {[astro-ph.CO]}

\bibitem{KAGRA:2013rdx}
B.P. Abbott, et~al., {Prospects for observing and localizing gravitational-wave
  transients with Advanced LIGO, Advanced Virgo and KAGRA}.
\newblock Living Rev. Rel. \textbf{19}, 1 (2016).
\newblock \doi{10.1007/s41114-020-00026-9}.
\newblock {\href{https://arxiv.org/abs/1304.0670}{{arXiv:1304.0670}}} {[gr-qc]}

\bibitem{LIGOScientific:2014pky}
J.~Aasi, et~al., {Advanced LIGO}.
\newblock Class. Quant. Grav. \textbf{32}, 074001 (2015).
\newblock \doi{10.1088/0264-9381/32/7/074001}.
\newblock {\href{https://arxiv.org/abs/1411.4547}{{arXiv:1411.4547}}} {[gr-qc]}

\bibitem{VIRGO:2014yos}
F.~Acernese, et~al., {Advanced Virgo: a second-generation interferometric
  gravitational wave detector}.
\newblock Class. Quant. Grav. \textbf{32}(2), 024001 (2015).
\newblock \doi{10.1088/0264-9381/32/2/024001}.
\newblock {\href{https://arxiv.org/abs/1408.3978}{{arXiv:1408.3978}}} {[gr-qc]}

\bibitem{KAGRA:2020tym}
T.~Akutsu, et~al., {Overview of KAGRA: Detector design and construction
  history}.
\newblock PTEP \textbf{2021}(5), 05A101 (2021).
\newblock \doi{10.1093/ptep/ptaa125}.
\newblock {\href{https://arxiv.org/abs/2005.05574}{{arXiv:2005.05574}}}
  {[physics.ins-det]}

\bibitem{LIGOScientific:2017adf}
B.P. Abbott, et~al., {A gravitational-wave standard siren measurement of the
  Hubble constant}.
\newblock Nature \textbf{551}(7678), 85--88 (2017).
\newblock \doi{10.1038/nature24471}.
\newblock {\href{https://arxiv.org/abs/1710.05835}{{arXiv:1710.05835}}}
  {[astro-ph.CO]}

\bibitem{LIGOScientific:2021aug}
R.~Abbott, et~al., {Constraints on the Cosmic Expansion History from
  GWTC\textendash{}3}.
\newblock Astrophys. J. \textbf{949}(2), 76 (2023).
\newblock \doi{10.3847/1538-4357/ac74bb}.
\newblock {\href{https://arxiv.org/abs/2111.03604}{{arXiv:2111.03604}}}
  {[astro-ph.CO]}

\bibitem{Ezquiaga:2022zkx}
J.M. Ezquiaga, D.E. Holz, {Spectral Sirens: Cosmology from the Full Mass
  Distribution of Compact Binaries}.
\newblock Phys. Rev. Lett. \textbf{129}(6), 061102 (2022).
\newblock \doi{10.1103/PhysRevLett.129.061102}.
\newblock {\href{https://arxiv.org/abs/2202.08240}{{arXiv:2202.08240}}}
  {[astro-ph.CO]}

\bibitem{Guidorzi:2017ogy}
C.~Guidorzi, et~al., {Improved Constraints on $H_0$ from a Combined Analysis of
  Gravitational-wave and Electromagnetic Emission from GW170817}.
\newblock Astrophys. J. Lett. \textbf{851}(2), L36 (2017).
\newblock \doi{10.3847/2041-8213/aaa009}.
\newblock {\href{https://arxiv.org/abs/1710.06426}{{arXiv:1710.06426}}}
  {[astro-ph.CO]}

\bibitem{Schutz:1986gp}
B.F. Schutz, {Determining the Hubble Constant from Gravitational Wave
  Observations}.
\newblock Nature \textbf{323}, 310--311 (1986).
\newblock \doi{10.1038/323310a0}

\bibitem{DelPozzo:2011vcw}
W.~Del~Pozzo, {Inference of the cosmological parameters from gravitational
  waves: application to second generation interferometers}.
\newblock Phys. Rev. D \textbf{86}, 043011 (2012).
\newblock \doi{10.1103/PhysRevD.86.043011}.
\newblock {\href{https://arxiv.org/abs/1108.1317}{{arXiv:1108.1317}}}
  {[astro-ph.CO]}

\bibitem{Gray:2019ksv}
R.~Gray, et~al., {Cosmological inference using gravitational wave standard
  sirens: A mock data analysis}.
\newblock Phys. Rev. D \textbf{101}(12), 122001 (2020).
\newblock \doi{10.1103/PhysRevD.101.122001}.
\newblock {\href{https://arxiv.org/abs/1908.06050}{{arXiv:1908.06050}}}
  {[gr-qc]}

\bibitem{Mukherjee:2020hyn}
S.~Mukherjee, B.D. Wandelt, S.M. Nissanke, A.~Silvestri, {Accurate precision
  Cosmology with redshift unknown gravitational wave sources}.
\newblock Phys. Rev. D \textbf{103}(4), 043520 (2021).
\newblock \doi{10.1103/PhysRevD.103.043520}.
\newblock {\href{https://arxiv.org/abs/2007.02943}{{arXiv:2007.02943}}}
  {[astro-ph.CO]}

\bibitem{Mukherjee:2022afz}
S.~Mukherjee, A.~Krolewski, B.D. Wandelt, J.~Silk, {Cross-correlating dark
  sirens and galaxies: constraints on $H_0$ from GWTC-3 of LIGO-Virgo-KAGRA}
  (2022).
\newblock {\href{https://arxiv.org/abs/2203.03643}{{arXiv:2203.03643}}}
  {[astro-ph.CO]}

\bibitem{Gray:2023wgj}
R.~Gray, et~al., {Joint cosmological and gravitational-wave population
  inference using dark sirens and galaxy catalogues}.
\newblock JCAP \textbf{12}, 023 (2023).
\newblock \doi{10.1088/1475-7516/2023/12/023}.
\newblock {\href{https://arxiv.org/abs/2308.02281}{{arXiv:2308.02281}}}
  {[astro-ph.CO]}

\bibitem{Farr:2019twy}
W.M. Farr, M.~Fishbach, J.~Ye, D.~Holz, {A Future Percent-Level Measurement of
  the Hubble Expansion at Redshift 0.8 With Advanced LIGO}.
\newblock Astrophys. J. Lett. \textbf{883}(2), L42 (2019).
\newblock \doi{10.3847/2041-8213/ab4284}.
\newblock {\href{https://arxiv.org/abs/1908.09084}{{arXiv:1908.09084}}}
  {[astro-ph.CO]}

\bibitem{You:2020wju}
Z.Q. You, X.J. Zhu, G.~Ashton, E.~Thrane, Z.H. Zhu, {Standard-siren cosmology
  using gravitational waves from binary black holes}.
\newblock Astrophys. J. \textbf{908}(2), 215 (2021).
\newblock \doi{10.3847/1538-4357/abd4d4}.
\newblock {\href{https://arxiv.org/abs/2004.00036}{{arXiv:2004.00036}}}
  {[astro-ph.CO]}

\bibitem{Mastrogiovanni:2021wsd}
S.~Mastrogiovanni, K.~Leyde, C.~Karathanasis, E.~Chassande-Mottin, D.A. Steer,
  J.~Gair, A.~Ghosh, R.~Gray, S.~Mukherjee, S.~Rinaldi, {On the importance of
  source population models for gravitational-wave cosmology}.
\newblock Phys. Rev. D \textbf{104}(6), 062009 (2021).
\newblock \doi{10.1103/PhysRevD.104.062009}.
\newblock {\href{https://arxiv.org/abs/2103.14663}{{arXiv:2103.14663}}}
  {[gr-qc]}

\bibitem{Karathanasis:2022rtr}
C.~Karathanasis, S.~Mukherjee, S.~Mastrogiovanni, {Binary black holes
  population and cosmology in new lights: signature of PISN mass and formation
  channel in GWTC-3}.
\newblock Mon. Not. Roy. Astron. Soc. \textbf{523}(3), 4539--4555 (2023).
\newblock \doi{10.1093/mnras/stad1373}.
\newblock {\href{https://arxiv.org/abs/2204.13495}{{arXiv:2204.13495}}}
  {[astro-ph.CO]}

\bibitem{Zevin:2017evb}
M.~Zevin, C.~Pankow, C.L. Rodriguez, L.~Sampson, E.~Chase, V.~Kalogera, F.A.
  Rasio, {Constraining Formation Models of Binary Black Holes with
  Gravitational-Wave Observations}.
\newblock Astrophys. J. \textbf{846}(1), 82 (2017).
\newblock \doi{10.3847/1538-4357/aa8408}.
\newblock {\href{https://arxiv.org/abs/1704.07379}{{arXiv:1704.07379}}}
  {[astro-ph.HE]}

\bibitem{Mapelli:2020vfa}
M.~Mapelli, {Binary Black Hole Mergers: Formation and Populations}.
\newblock Front. Astron. Space Sci. \textbf{7}, 38 (2020).
\newblock \doi{10.3389/fspas.2020.00038}.
\newblock {\href{https://arxiv.org/abs/2105.12455}{{arXiv:2105.12455}}}
  {[astro-ph.HE]}

\bibitem{Zevin:2020gbd}
M.~Zevin, S.S. Bavera, C.P.L. Berry, V.~Kalogera, T.~Fragos, P.~Marchant, C.L.
  Rodriguez, F.~Antonini, D.E. Holz, C.~Pankow, {One Channel to Rule Them All?
  Constraining the Origins of Binary Black Holes Using Multiple Formation
  Pathways}.
\newblock Astrophys. J. \textbf{910}(2), 152 (2021).
\newblock \doi{10.3847/1538-4357/abe40e}.
\newblock {\href{https://arxiv.org/abs/2011.10057}{{arXiv:2011.10057}}}
  {[astro-ph.HE]}

\bibitem{Mandel:2018hfr}
I.~Mandel, A.~Farmer, {Merging stellar-mass binary black holes}.
\newblock Phys. Rept. \textbf{955}, 1--24 (2022).
\newblock \doi{10.1016/j.physrep.2022.01.003}.
\newblock {\href{https://arxiv.org/abs/1806.05820}{{arXiv:1806.05820}}}
  {[astro-ph.HE]}

\bibitem{Marchant:2023wno}
P.~Marchant, J.~Bodensteiner, {The Evolution of Massive Binary Stars}  (2023).
\newblock {\href{https://arxiv.org/abs/2311.01865}{{arXiv:2311.01865}}}
  {[astro-ph.SR]}

\bibitem{Mukherjee:2021rtw}
S.~Mukherjee, {The redshift dependence of black hole mass distribution: is it
  reliable for standard sirens cosmology?}
\newblock Mon. Not. Roy. Astron. Soc. \textbf{515}(4), 5495--5505 (2022).
\newblock \doi{10.1093/mnras/stac2152}.
\newblock {\href{https://arxiv.org/abs/2112.10256}{{arXiv:2112.10256}}}
  {[astro-ph.CO]}

\bibitem{Pierra:2023deu}
G.~Pierra, S.~Mastrogiovanni, S.~Perri\`es, M.~Mapelli, {Study of systematics
  on the cosmological inference of the Hubble constant from gravitational wave
  standard sirens}.
\newblock Phys. Rev. D \textbf{109}(8), 083504 (2024).
\newblock \doi{10.1103/PhysRevD.109.083504}.
\newblock {\href{https://arxiv.org/abs/2312.11627}{{arXiv:2312.11627}}}
  {[astro-ph.CO]}

\bibitem{LIGOScientific:2020kqk}
R.~Abbott, et~al., {Population Properties of Compact Objects from the Second
  LIGO-Virgo Gravitational-Wave Transient Catalog}.
\newblock Astrophys. J. Lett. \textbf{913}(1), L7 (2021).
\newblock \doi{10.3847/2041-8213/abe949}.
\newblock {\href{https://arxiv.org/abs/2010.14533}{{arXiv:2010.14533}}}
  {[astro-ph.HE]}

\bibitem{KAGRA:2021duu}
R.~Abbott, et~al., {Population of Merging Compact Binaries Inferred Using
  Gravitational Waves through GWTC-3}.
\newblock Phys. Rev. X \textbf{13}(1), 011048 (2023).
\newblock \doi{10.1103/PhysRevX.13.011048}.
\newblock {\href{https://arxiv.org/abs/2111.03634}{{arXiv:2111.03634}}}
  {[astro-ph.HE]}

\bibitem{Rinaldi:2021bhm}
S.~Rinaldi, W.~Del~Pozzo, {(H)DPGMM: a hierarchy of Dirichlet process Gaussian
  mixture models for the inference of the black hole mass function}.
\newblock Mon. Not. Roy. Astron. Soc. \textbf{509}(4), 5454--5466 (2021).
\newblock \doi{10.1093/mnras/stab3224}.
\newblock {\href{https://arxiv.org/abs/2109.05960}{{arXiv:2109.05960}}}
  {[astro-ph.IM]}

\bibitem{Mandel:2016prl}
I.~Mandel, W.M. Farr, A.~Colonna, S.~Stevenson, P.~Ti\v{n}o, J.~Veitch,
  {Model-independent inference on compact-binary observations}.
\newblock Mon. Not. Roy. Astron. Soc. \textbf{465}(3), 3254--3260 (2017).
\newblock \doi{10.1093/mnras/stw2883}.
\newblock {\href{https://arxiv.org/abs/1608.08223}{{arXiv:1608.08223}}}
  {[astro-ph.HE]}

\bibitem{Mandel:2018mve}
I.~Mandel, W.M. Farr, J.R. Gair, {Extracting distribution parameters from
  multiple uncertain observations with selection biases}.
\newblock Mon. Not. Roy. Astron. Soc. \textbf{486}(1), 1086--1093 (2019).
\newblock \doi{10.1093/mnras/stz896}.
\newblock {\href{https://arxiv.org/abs/1809.02063}{{arXiv:1809.02063}}}
  {[physics.data-an]}

\bibitem{Tiwari:2020vym}
V.~Tiwari, {VAMANA: modeling binary black hole population with minimal
  assumptions}.
\newblock Class. Quant. Grav. \textbf{38}(15), 155007 (2021).
\newblock \doi{10.1088/1361-6382/ac0b54}.
\newblock {\href{https://arxiv.org/abs/2006.15047}{{arXiv:2006.15047}}}
  {[astro-ph.HE]}

\bibitem{Edelman:2021zkw}
B.~Edelman, Z.~Doctor, J.~Godfrey, B.~Farr, {Ain\textquoteright{}t No Mountain
  High Enough: Semiparametric Modeling of
  LIGO\textendash{}Virgo\textquoteright{}s Binary Black Hole Mass
  Distribution}.
\newblock Astrophys. J. \textbf{924}(2), 101 (2022).
\newblock \doi{10.3847/1538-4357/ac3667}.
\newblock {\href{https://arxiv.org/abs/2109.06137}{{arXiv:2109.06137}}}
  {[astro-ph.HE]}

\bibitem{Sadiq:2021fin}
J.~Sadiq, T.~Dent, D.~Wysocki, {Flexible and fast estimation of binary merger
  population distributions with an adaptive kernel density estimator}.
\newblock Phys. Rev. D \textbf{105}(12), 123014 (2022).
\newblock \doi{10.1103/PhysRevD.105.123014}.
\newblock {\href{https://arxiv.org/abs/2112.12659}{{arXiv:2112.12659}}}
  {[gr-qc]}

\bibitem{Edelman:2022ydv}
B.~Edelman, B.~Farr, Z.~Doctor, {Cover Your Basis: Comprehensive Data-driven
  Characterization of the Binary Black Hole Population}.
\newblock Astrophys. J. \textbf{946}(1), 16 (2023).
\newblock \doi{10.3847/1538-4357/acb5ed}.
\newblock {\href{https://arxiv.org/abs/2210.12834}{{arXiv:2210.12834}}}
  {[astro-ph.HE]}

\bibitem{Callister:2023tgi}
T.A. Callister, W.M. Farr, {Parameter-Free Tour of the Binary Black Hole
  Population}.
\newblock Phys. Rev. X \textbf{14}(2), 021005 (2024).
\newblock \doi{10.1103/PhysRevX.14.021005}.
\newblock {\href{https://arxiv.org/abs/2302.07289}{{arXiv:2302.07289}}}
  {[astro-ph.HE]}

\bibitem{Ray:2023upk}
A.~Ray, I.~Maga\~na Hernandez, S.~Mohite, J.~Creighton, S.~Kapadia,
  {Nonparametric Inference of the Population of Compact Binaries from
  Gravitational-wave Observations Using Binned Gaussian Processes}.
\newblock Astrophys. J. \textbf{957}(1), 37 (2023).
\newblock \doi{10.3847/1538-4357/acf452}.
\newblock {\href{https://arxiv.org/abs/2304.08046}{{arXiv:2304.08046}}}
  {[gr-qc]}

\bibitem{Li:2023yyt}
Y.J. Li, Y.Z. Wang, S.P. Tang, Y.Z. Fan, {Resolving the Stellar-Collapse and
  Hierarchical-Merger Origins of the Coalescing Black Holes}.
\newblock Phys. Rev. Lett. \textbf{133}(5), 051401 (2024).
\newblock \doi{10.1103/PhysRevLett.133.051401}.
\newblock {\href{https://arxiv.org/abs/2303.02973}{{arXiv:2303.02973}}}
  {[astro-ph.HE]}

\bibitem{Farah:2024xub}
A.M. Farah, T.A. Callister, J.M. Ezquiaga, M.~Zevin, D.E. Holz, {No need to
  know: astrophysics-free gravitational-wave cosmology}  (2024).
\newblock {\href{https://arxiv.org/abs/2404.02210}{{arXiv:2404.02210}}}
  {[astro-ph.CO]}

\bibitem{MaganaHernandez:2024uty}
I.~Maga\~na Hernandez, A.~Ray, {Beyond Gaps and Bumps: Spectral Siren Cosmology
  with Non-Parametric Population Models}  (2024).
\newblock {\href{https://arxiv.org/abs/2404.02522}{{arXiv:2404.02522}}}
  {[astro-ph.CO]}

\bibitem{Li:2024rmi}
Y.J. Li, S.P. Tang, Y.Z. Wang, Y.Z. Fan, {Multi-spectral sirens:
  Gravitational-wave cosmology with (multi-) sub-populations of binary black
  holes}  (2024).
\newblock {\href{https://arxiv.org/abs/2406.11607}{{arXiv:2406.11607}}}
  {[astro-ph.CO]}

\bibitem{Rinaldi:2022kyg}
S.~Rinaldi, W.~Del~Pozzo, {Rapid localization of gravitational wave hosts with
  FIGARO}.
\newblock Mon. Not. Roy. Astron. Soc. \textbf{517}(1), L5--L10 (2022).
\newblock \doi{10.1093/mnrasl/slac101}.
\newblock {\href{https://arxiv.org/abs/2205.07252}{{arXiv:2205.07252}}}
  {[astro-ph.IM]}

\bibitem{Rinaldi:2024eep}
S.~Rinaldi, W.~Del~Pozzo, {FIGARO: hierarchical non-parametric inference for
  population studies}.
\newblock J. Open Source Softw. \textbf{9}(97), 6589 (2024).
\newblock \doi{10.21105/joss.06589}

\bibitem{Essick:2023upv}
R.~Essick, M.~Fishbach, {Ensuring Consistency between Noise and Detection in
  Hierarchical Bayesian Inference}.
\newblock Astrophys. J. \textbf{962}(2), 169 (2024).
\newblock \doi{10.3847/1538-4357/ad1604}.
\newblock {\href{https://arxiv.org/abs/2310.02017}{{arXiv:2310.02017}}}
  {[gr-qc]}

\bibitem{2020SciPy-NMeth}
P.~Virtanen, R.~Gommers, T.E. Oliphant, M.~Haberland, T.~Reddy, D.~Cournapeau,
  E.~Burovski, P.~Peterson, W.~Weckesser, J.~Bright, S.J. {van der Walt},
  M.~Brett, J.~Wilson, K.J. Millman, N.~Mayorov, A.R.J. Nelson, E.~Jones,
  R.~Kern, E.~Larson, C.J. Carey, {\.I}.~Polat, Y.~Feng, E.W. Moore,
  J.~{VanderPlas}, D.~Laxalde, J.~Perktold, R.~Cimrman, I.~Henriksen, E.A.
  Quintero, C.R. Harris, A.M. Archibald, A.H. Ribeiro, F.~Pedregosa, P.~{van
  Mulbregt}, {SciPy 1.0 Contributors}, {{SciPy} 1.0: Fundamental Algorithms for
  Scientific Computing in Python}.
\newblock Nature Methods \textbf{17}, 261--272 (2020).
\newblock \doi{10.1038/s41592-019-0686-2}

\bibitem{ng_2024_13968239}
T.~Ng, S.~Rinaldi, H.~Otto.
\newblock {Inferring cosmology from gravitational waves using non-parametric
  detector-frame mass distribution: Data Release} (2024).
\newblock \doi{10.5281/zenodo.13968239}.
\newblock \urlprefix\url{https://doi.org/10.5281/zenodo.13968239}

\bibitem{Lorenzo-Medina:2024opt}
A.~Lorenzo-Medina, T.~Dent, {A physically modelled selection function for
  compact binary mergers in the LIGO-Virgo O3 run and beyond}  (2024).
\newblock {\href{https://arxiv.org/abs/2408.13383}{{arXiv:2408.13383}}}
  {[gr-qc]}

\bibitem{LIGOScientific:2021usb}
R.~Abbott, et~al., {GWTC-2.1: Deep extended catalog of compact binary
  coalescences observed by LIGO and Virgo during the first half of the third
  observing run}.
\newblock Phys. Rev. D \textbf{109}(2), 022001 (2024).
\newblock \doi{10.1103/PhysRevD.109.022001}.
\newblock {\href{https://arxiv.org/abs/2108.01045}{{arXiv:2108.01045}}}
  {[gr-qc]}

\bibitem{ligo_scientific_collaboration_and_virgo_2022_6513631}
L.S. Collaboration, V.~Collaboration.
\newblock {GWTC-2.1: Deep Extended Catalog of Compact Binary Coalescences
  Observed by LIGO and Virgo During the First Half of the Third Observing Run -
  Parameter Estimation Data Release} (2022).
\newblock \doi{10.5281/zenodo.6513631}.
\newblock \urlprefix\url{https://doi.org/10.5281/zenodo.6513631}

\bibitem{KAGRA:2021vkt}
R.~Abbott, et~al., {GWTC-3: Compact Binary Coalescences Observed by LIGO and
  Virgo during the Second Part of the Third Observing Run}.
\newblock Phys. Rev. X \textbf{13}(4), 041039 (2023).
\newblock \doi{10.1103/PhysRevX.13.041039}.
\newblock {\href{https://arxiv.org/abs/2111.03606}{{arXiv:2111.03606}}}
  {[gr-qc]}

\bibitem{ligo_scientific_collaboration_and_virgo_2023_8177023}
L.S. Collaboration, V.~Collaboration, K.~Collaboration.
\newblock {GWTC-3: Compact Binary Coalescences Observed by LIGO and Virgo
  During the Second Part of the Third Observing Run — Parameter estimation
  data release} (2023).
\newblock \doi{10.5281/zenodo.8177023}.
\newblock \urlprefix\url{https://doi.org/10.5281/zenodo.8177023}

\bibitem{LIGOScientific:2019lzm}
R.~Abbott, et~al., {Open data from the first and second observing runs of
  Advanced LIGO and Advanced Virgo}.
\newblock SoftwareX \textbf{13}, 100658 (2021).
\newblock \doi{10.1016/j.softx.2021.100658}.
\newblock {\href{https://arxiv.org/abs/1912.11716}{{arXiv:1912.11716}}}
  {[gr-qc]}

\bibitem{KAGRA:2023pio}
R.~Abbott, et~al., {Open Data from the Third Observing Run of LIGO, Virgo,
  KAGRA, and GEO}.
\newblock Astrophys. J. Suppl. \textbf{267}(2), 29 (2023).
\newblock \doi{10.3847/1538-4365/acdc9f}.
\newblock {\href{https://arxiv.org/abs/2302.03676}{{arXiv:2302.03676}}}
  {[gr-qc]}

\bibitem{Luger2021}
R.~{Luger}, M.~{Bedell}, D.~{Foreman-Mackey}, I.J.M. {Crossfield}, L.L. {Zhao},
  D.W. {Hogg}, {Mapping stellar surfaces III: An Efficient, Scalable, and
  Open-Source Doppler Imaging Model}.
\newblock arXiv e-prints arXiv:2110.06271 (2021).
\newblock {\href{https://arxiv.org/abs/2110.06271}{{arXiv:2110.06271}}}
  {[astro-ph.SR]}

\end{thebibliography}

\end{document}